
\magnification=1200
\baselineskip=18pt
\null
\hfill TAUP N237-94
\bigskip
\centerline{\bf Cosmological Surrealism: More than
``Eternal Reality" is Needed}
\bigskip
\bigskip
\centerline{by}
\bigskip
\centerline{Yuval Ne'eman$^{*\#\&}$}
\centerline{Raymond and Beverly Sackler Faculty of Exact Sciences}
\centerline{Tel-Aviv University, Tel-Aviv, Israel}
\vskip 3 true cm
\bigskip
\centerline{\bf Abstract}

Inflationary Cosmology makes the universe ``eternal" and provides for
recurrent
universe creation, ad infinitum -- making it also plausible to assume
that
``our" Big Bang was also preceeded by others, etc.. However, GR tells
us that
in the ``parent" universe's reference frame, the newborn universe's
expansion
will never start. Our picture of ``reality" in spacetime has to be
enlarged.
\vskip 4 true cm
\medskip
\item{$*$}
Wolfson Distinguished Chair in Theoretical Physics.
\item{$\#$}
Also on leave from the University of Texas, Austin, Center for
Particle
Physics.
\item{$\&$}
 Royal Society - Israel National Academy of Sciences Visiting
Professor,
  Institute of Astronomy at the University of Cambridge, Cambridge,
UK
  and Physics Department, Imperial College of Science, Technology and
  Medicine, London, UK.
\vfill\eject

One is used to associate the Foundations of Quantum Mechanics with
fundamental metaphysical issues, such as EPR's ``is there an
underlying
reality?" [1]. I would like to suggest that as a result of the recent
advances in cosmological studies, in the context of the Inflationary
Model [2,3], physics has effectively undergone yet another, perhaps
its
most profound revolution. This is conceptually comparable to the 1905
rejection of absolute time (the negation of absolute space was
conceptually
natural, as noted by Bishop Berkeley, by Newton himself and
especially by
Mach; the revolutionary aspect was limited in that case to the
rejection of
the Newtonian formalism).

Yet another comparable conceptual transformation happened with
Aspect's
(experimentally derived [4]) negative answer to EPR's above-mentioned
querry, as explicited by Bell's inequalities [5]. Here, one's
intuitively perceived ``objective" material reality is now replaced
by just
potentially materializable, but otherwise ``ethereal" amplitudes. In
fact,
the latest revolution, which we are noting and delineating in this
letter, does
bear some resemblance to De Witt's version [6]) of Everett's [7]
``Many-Worlds"
interpretation of Quantum Mechanics; except that the latter example
should not
-- in our view -- be considered as final, since a literal
interpretation might
yet be dispensed with, should a proper mechanism for ``dicing" be
developed
(following Einstein's ``God doesn't play dice"). We have conjectured
that a
mechanism in the nature of a spontaneous symmetry breakdown [8,9]
might bring
about the materialization of one component of the state-vector and
the
cancellation of the rest, thus avoiding the need for the ``many
worlds" to
exist.

The present observation is the following:

a) In the creation of black holes through gravitational collapse (a
topic
first investigated after the discovery of quasars and the conjecture
that
their energy is supplied by gravitational collapse of very massive
stars [10]),
the collapsing matter will never reach its Schwarzschild radius [11],
in the
reference frame of a distant outside observer A [12]. However, in the
reference
frame of an observer B, sitting on the collapsing star and falling
into the
black hole, the Schwarzschild radius is reached and crossed within
hours or
minutes from the collapse's start; the unfortunate B is ``eaten up"
by the
$r=0$ singularity after a comparable stretch of (his) time. This is
best
studied in Kruskal-Szekeres coordinates [13]. Since the Principle of
Covariance
denies the existence of any ``preferred" reference frame, the ``post-
future"
(i.e. that which comes {\it after A's future}, which is also
``our's") ``last
trip" of B already contains the seeds of our announced metaphysical
revolution:
where (and ``when") indeed will A (or the outside) ``be", when B is
half-way
between the Schwarzschild radius and $r=0$? Or alternatively, how can
B be
allowed his (or her) reference frame, in the equalitarian regime of
Covariance,
if we can claim in all finality that B {\it will never cross that
Schwarzschild radius, in our spacetime reality?}. Before the
emergence of
Inflationary Cosmology, however, B could be dismissed as some kind of
thin
``fringe" on the borders of reality -- ``an extra half-hour" added to
Eternity,
perhaps an oddity of our description of spacetime. And yet, that half-
hour
somehow does not overlap with ${\it our}$ reality? Are there perhaps
other
``realities"? Can we accept more than one reality, just as there are
any number
of reference frames?

b) A similar situation arises in ``Eternal" Inflationary Cosmology
[3]. New
universes can be created (e.g. [14,15]) through a mechanism
(Inflation) which
emulates the de Sitter Model [16] in the first $10^{-35}$ sec., then
``exits"
this mode and settles in a flat $k=0$ Friedmannian quasi-linear
expansion.
The first (inflationary) phase can be induced whenever a vacuum
fluctuation,
or some other mechanism, e.g. a collision between two $10^{11} GeV$
cosmic
rays [17], might generate, in a very tiny spatial region, an energy-
density
larger than $10^{75} g/cm^{3}$, i.e. about $10^{14} - 10^{15} GeV$,
contained
in a volume whose linear dimensions are of the order of the
corresponding
Compton wavelength. The tiny system might then ``settle" for a while
as a
``false vacuum" in that state: in an unstable (symmetric?) solution
of the
(otherwise) spontaneously-broken-symmetry mechanism of a GUT,
provided it
would have gotten there through supercooling, for instance, or some
alternative non-turbulent phase; this would correspond to having a
region with
a cosmological constant $\lambda$,  the classical GRG representation
of the
quantum vacuum energy. It would then trigger a de Sitter exponential
expansion
$S = exp (Ht)$, with Hubble constant
$H = \sqrt{(8\pi G \lambda)\over 3 c^{2}}$.

Outside observers A will just note the creation of a tiny black-hole,
a
Schwarzschild solution as in (a) above, with only the very beginnings
of an
expansion, lasting in this state ``forever", i.e. with $t \rightarrow
\infty$.
One way of visualizing this phase is to remember that the exponential
growth
of the tiny region is like a very fast ``unfurling" of huge amounts
of new
space, i.e. the larger parts of the original de Sitter new universe
are
infinitely red-shifted with respect to A. Our entire universe is in
an A-type
frame and will {\it never} see the transformation of that tiny false
vacuum
region into anything else. However, for an inside frame of reference
B, we
have the birth of a de Sitter universe, a Big Bang, followed by the
exit phase,
then evolving into a new Friedmann (flat) universe -- and perhaps,
some
$10^{10}$ years later, physicists discussing concepts of reality. The
B
picture is best studied in Gibbons-Hawking coordinates [18]. The new
universe
might have involved a singularity (a time-like half-line) due to the
Penrose
theorem -- except that quantum tunneling makes it possible, for that
new
universe, to avoid the singularity. In one such solution [14], the
new cosmos
starts with a total mass smaller than some critical value.
Classically, it
would then recollapse without inflation and would also not come under
Penrose's
conditions for the singularity to occur. Instead, however, it then
quantum-tunnels into the exponentially inflating solution (occuring
only for
masses larger than the critical value, classically) and goes on to
make a
universe, having thus managed to skip the singular stage.

As a result, the new universe carries no singularity blemish and is
no
different from its parent, ``our" present universe. Presumably, this
is also
how the universe we live in came into being, with an {\it eternal}
lifetime
and with no singularities, neither in its past nor in its future. We
should
thus extend the Principle of Covariance to all such universes. They
are all
eternal - except that this is meaningless within our present
conceptual
framework: the new universe will never exist in our frame A, in all
our
time; and yet it is as good as our own universe, will have (in its B
frame)
galaxies and suns and perhaps physicists. So where and when does it
exist?
Never, says A. Forever, says B. Note that the two did overlap {\it
before the
``happy event" which triggered the birth of a universe}, out of a
given false
vacuum in a region of ``our" universe. They then separate, B going it
by
itself, observing A fading flashing out its eternity in the infinitely
redshifted environment of the new Big Bang..

Clearly, ``eternity", as mentioned in Inflationary Cosmology [3] is
not an
adequate answer -- it just relates to A, to the eternity of ``our"
reality.
There is, however, (perhaps) a countable infinity of such
``eternities",
branching out from each other, then separating, with the offspring
``hibernating" and never being born, in the parent universe's
reality, to ``the
end of time" = our eternity. And yet, beyond this eternity, there is
another
full-fledged universe, the offspring, flourishing and ``realizing
itself".
Clearly, this new picture calls for our conceptual framework to admit
``surrealism", {\it i.e. ``existence" beyond space and time as we
know them.}

The theoretical basis for this conceptual jump has been around since
the
earliest beginnings of General Relativity, since all it involves is
the
Schwarzschild solution [11] and the de Sitter Model [16], perhaps
also the
Einstein-Rosen Bridge [19]. Interestingly enough, we came close to
such a
picture in our lagging-core hypothesis for the quasars [20], except
that
that quasar interpretation required all these de Sitter solution
quasars
to emerge into {\it the same universe}, no trivial requirement. The
issue
does exist for collapsing black holes, but these could be disregarded
as far
as their B picture was concerned, by regarding them as ``odd" pieces
of our
reality, exceptional covariant frames never realizing their full
physical
content. This position can no more be justified in an ever
multiplying
Inflationary Cosmology, in which one of the main points is the
physical
non-uniqueness of universe creation, yet another sur-grandiose
Copernican
rejection of ``our" centrality. We thus have to learn to enlarge our
conception of what ``is" beyond our space and time. This is sur-
history and
surreality..

I would like to thank Prof. D. Lynden-Bell and the Institute of
Astronomy
at the University of Cambridge for the Institute's hospitality during
the
fall trimester of 1993; it was in the inspiring creative atmosphere
of the
Institute and having absorbed some of the lessons derived by Lynden-
Bell
and J. Katz, relating to black holes, etc., that these ideas first
started
forming. I would also like to thank Prof. C. Isham for the
hospitality of
Imperial College and for interesting discussions relating to Quantum
Gravity.
\vfill\eject
\noindent
{\bf References.}
\bigskip
\item{1.}
Einstein A., Podolsky B. and Rosen N., 1935, {\it Phys. Rev.} {\bf
48}, 777.
\item{2.}  Guth A.H., 1981, {\it Phys. Rev.} {\bf D23}, 347;
    Linde A., 1982, {\it Phys. Lett.} {\bf B108}, 389;
    Albrecht A., and Steinhardt P.J., (1982) {\it Phys. Rev. Lett.},
{\bf 48}
    1220.
    La D. and Steinhardt P.J., 1989 {\it Phys. Rev. Lett.}, {\bf 62}
376.
\item{3.} Recent reviews:
    Guth A.H., 1993 {\it Proc. Nat. Acad. Sci. USA}, {\bf 90} 4871;
    Linde A., 1991, {\it Gravitation and Modern Cosmology} 1991,
Zichichi A.,
    ed., (New York: Plenum Press);
    Steinhardt P.J., 1993, {\it Class. Quantum Grav.}, {\bf 10} S33.
\item{4.}
Aspect A., Grangier P. and Roger, G., 1982 {\it Phys. Rev. Lett.}
{\bf 49}
    91; Aspect A., Dalibard J. and Roger G., {\it idem} {\bf 49} 1804.
\item{5.}
Bell J.S., 1966 {\it Rev. Mod. Phys.} {\bf 38} 447.
\item{6.} De Witt B.S., 1968 {\it Batelle Rencontres}I,
C. De Witt and J.A. Wheeler eds., (NY: Benjamin)
\item{7.} Everett H. III, 1957 {\it Rev. Mod. Phys.} {\bf 29}, 454.
\item{8.}
  Ne'eman Y., 1988, {\it Microphysical Reality and Quantum
Formalism},
    Selleri F. ed., (Amsterdam: Kluwer) p. 141.
    Ne'eman Y., {\it Decoherence Plus Spontaneous Symmetry Breakdown
Generate
    the Ohmic View of the State-Vector Collapse}, to be pub. in Proc.
1993
    Cologne Symp. {\it Foundations of Modern Physics}.
\item{9.}
Ghirardi G.C., Rimini A. and Weber T., 1986 {\it Phys. Rev.} {\bf
D34} 470
    and {\it idem} {\bf 36} 3287.
\item{10.}
Hoyle F., Fowler W.A., Burbidge G.R. and Burbidge E.M., 1964 {\it Ap.
J.}
    {\bf 139} 909.
\item{11.}
Schwarzschild K., 1916 {\it Sitzber. Deut. Akad. Wiss. Berlin, Kl.
Math-
    Phys. Tech.}, 189.
\item{12.}
A beautiful illustration of this state of affairs is provided in
Frederick
    Pohl's novel, ``Beyond the Blue Horizon". The hero suffers for
thirty years
    from a depression caused by his realization that throughout his
entire
    lifetime, his fiancee is suffering and shocked by his own
behavior. She is
    in a spaceship falling into a black hole, which they had been
exploring
    together, each in his own spaceship. A fatal mistake on his part
caused
    her ship to be shoved into the black hole, while his ship thereby
recoiled
    and made it to safety. His entire lifetime therefore coincides
with one
    second of her time, just that second in which she is wondering
why he has
    abandoned her and perhaps even suspects his motives. She is
finally
    extracted from the hole and is now very much younger than her
lover of the
    previous second.
\item{13.}
Kruskal M.D., 1960 {\it Phys. Rev.} {\bf 119} 1743;
    Szekeres G., 1960 {\it Pub. Math. Debrecen} {\bf 7} 285.
\item{14.}
Farhi E., Guth A.H. and Guven J., 1990 {\it Nucl. Phys.} {\bf B339}
417.
\item{15.}
Fischler W., Morgan D. and Polchinski J., 1990 {\it Phys. Rev.} {\bf
D42}
    4042.
\item{16.}
de Sitter W., 1917 {\it Proc. Kon. Ned. Akad. v. Wetensch.}, {\bf 19}
1217.
\item{17.}
Hut P. and Rees M., 1983 {\it Nature}, {\bf 302} 508.
\item{18.}
Gibbons G.W. and Hawking S., 1977 {\it Phys. Rev.} {\bf D15} 2738.
\item{19.}
Einstein A., and Rosen N., 1935 {\it Phys. Rev.} {\bf 48} 73.
\item{20.}
Novikov I.D., 1964 {\it Astr. Zh.}, {\bf 41} 1075;
    Ne'eman Y., 1965 {\it Ap. J.}, {\bf 141} 1303;
    Ne'eman Y. and Tauber G., {\it Ap. J.}, {\bf 150} 755.
\end